\documentstyle[epsf,rotate,psfig]{mn}
\def\gsim{~\rlap{$>$}{\lower 1.0ex\hbox{$\sim$}}}

\title{ Spectroscopy of Globular Cluster Candidates in
the Sculptor Group Galaxies NGC~253 and NGC~55}

\author[M.A. Beasley \& R.M. Sharples]
	{M.A. Beasley \& R.M. Sharples \\
	     Department of Physics, University of Durham, Durham DH1 3LE}

\date{Accepted~~~~~~~~~~.   Received~~~~~~~~~~.}

\pubyear{1999}

\begin{document}

\maketitle

\begin{abstract}
We have obtained spectra for 103 published globular cluster candidates in 
the Sculptor Group galaxies NGC~253 and NGC~55. On the basis of radial 
velocities and digitized plate images, 14 globular clusters are identified in 
NGC~253 and one probable globular cluster is identified in NGC~55. 
The majority of the objects in the sample appear to be background galaxies. 
We have obtained and analysed COSMOS plate scans of NGC~253 and NGC~55 and use 
these along with the spectroscopically identified clusters to define new 
samples of globular cluster candidates in the two galaxies which should have 
reduced contamination.    
\end{abstract}

\begin{keywords}
galaxies: individual:NGC~253 -- galaxies: individual:NGC~55 -- galaxies: 
spiral -- galaxies: star clusters
\end{keywords}

\alph{footnote}
\section{Introduction}

The search for globular clusters in external galaxies
has progressed to the point where globular cluster systems have
now been studied in over 100 galaxies (e.g. Harris 1999).
In most cases these globular cluster systems
are identified as a statistical excess of images above
background, clustered around the parent galaxy. If the known
globular clusters in the Galaxy are assumed to be representative
of old cluster populations in external galaxies, then only
in the nearest galaxies are the brightest globular
clusters expected to be resolved in ground--based
images (see Harris et al. 1984).
The Sculptor group of galaxies forms a loose physical
association of about 15 members (de Vaucouleurs 1959, 1978). 
At a distance of 2.5~Mpc \cite{graham} it is generally believed to be 
the nearest aggregate of galaxies to our own Local Group 
\cite{devaucouleurs75}. 
Despite its relative proximity, very little is known of the globular 
cluster systems surrounding the major galaxies in the Sculptor group
(NGC~45, 55, 247, 253, 300 and 7793). Da Costa \& Graham (1982) obtained 
spectroscopy of three resolved cluster candidates in the field of NGC~55, 
an SB(s)m galaxy with $M_B = -19.7$
 \cite{devaucouleurs91}, and found all three to have velocities which 
agreed with that of NGC~55 itself.
A visual search for globular cluster candidates
around NGC~55 and NGC~253
has been made by Liller \& Alcaino (1983a,b)
using plates from the ESO 3.6 m telescope. These
authors found a total of 114 slightly diffuse objects
with the $B-V$ colours, magnitudes and sizes appropriate to those
of a globular cluster population similar to that in the Galaxy.
A more quantitative selection of candidates was later derived by
Blecha (1986) using profile analysis of images from the
Danish 1.5 m telescope.

The true nature of these globular cluster candidates, however,
can only be established via spectroscopy to determine accurate
radial velocities. Since these objects are marginally resolved, there
is not the same level of confusion with Galactic stars as is the case with
more distant cluster systems. Moreover, because the 
parent galaxies have low systemic velocities (${ V_{\rm N55}}$=129~kms$^{-1}$;
$V_{\rm N253}$ = 245~kms$^{-1}$) \cite{dacosta91}, there is little 
uncertainty in identifying background galaxies (see $\S 4$). In this paper
we present a spectroscopic survey of the Liller \& Alcaino (1983a,b)
and Blecha (1986) samples. The plan of the paper is as follows:
$\S 2$ contains the observations and data reduction, with the
radial velocity analysis presented in $\S 3$. In $\S 4$ we discuss the 
spectroscopically confirmed clusters and use this information together with 
COSMOS digitised plate scans to define a new sample of globular cluster 
candidates in $\S 5$. Conclusions are summarized in $\S 6$.

\section{Observations and Data Reduction}

\subsection{Sample Selection}

The sample was selected from the
list of globular clusters candidates 
published by Liller \& Alcaino (1983a) for NGC~55, and
from Liller \& Alcaino (1983b) and Blecha (1986)
for NGC~253. The candidates were taken within the 
radial limits $0<R<20$ arcmin from the centres of each  galaxy
and magnitude limits $18<B<20.5$.
For completeness, our candidates included twenty extra objects, six
labelled 'bright' and fourteen labelled 'blue' in the 
Liller \& Alcaino (1983a,b) papers. 
Accurate astrometric positions ($\pm$ 0.3 arcsec) for
each target were obtained using
a PDS measuring machine and reference stars from the 
Perth70 catalogue. At this stage any obvious galaxies
(usually low surface brightness objects showing
spiral structure) were expunged from the lists.
The final list of candidates contained 57 objects in NGC~55 and
58 in NGC~253.

\subsection{Observations}

The observations were obtained with the 3.9 m Anglo--Australian
Telescope (AAT) and the AUTOFIB fibre
positioner \cite{parry} to
obtain intermediate dispersion spectra
of up to 64 objects within a 40 arcminute diameter field.
Several fibres were used to monitor the sky background
spectrum and the faint background light from the parent galaxy halo.
Using a 600 lines mm$^{-1}$ grating in first order with the RGO 
spectrograph, we obtained spectra covering the range 3850--5700~\AA\
at a resolution of 4~\AA\ using 
the Image Photon Counting System (IPCS)
as the detector. For each galaxy $5\times 3000$~second exposures
were interspersed with 200~second exposures of a Cu--Ar--He calibration
lamp and 300~second exposures of blank sky regions.
Most of the observations
were obtained in rather poor seeing conditions (2--3 arcsec) so the final 
spectra have S/N ratios typically in the range 4--20.

\subsection{Data Reduction}

The data have been primarily reduced using the FIGARO data reduction package
with standard techniques to extract the individual spectra from the data 
frame and wavelength calibrate them using the exposures of the Cu--Ar--He
hollow cathode lamp. The rms residuals of the wavelength
calibration were typically 0.2~\AA.
Sky subtraction was based on the dedicated sky fibres
in each frame, with the fibre--to--fibre transmission variations
and vignetting along the spectrograph slit being removed using the 
blank sky exposures. In addition, we have corrected
for the fibre--to--fibre {\em spectral} response (most of which is
introduced when the spectra are extracted) using exposures of the
twilight sky spectrum whose shape is assumed to be constant across the field.  
During the run, exposures were obtained of several bright radial velocity 
standards by offsetting the stars into individual fibres. 
The spectra consisted of 925 channels at 2~\AA\ pixel$^{-1}$ and prior to 
cross--correlation analysis were rebinned onto a logarithmic wavelength scale 
with a velocity step of 127~kms$^{-1}$ per bin.

Table \ref{tab:templates} summarizes the template objects observed for the 
two galaxies.

\begin{table}
 \centering
  \caption{Template objects observed for NGC~55$^1$ and NGC~253$^2$ }
 \label{tab:templates} 
\begin{tabular}{lll}
\hline
   
Template & Object Type & Spectral Type \\
\hline
HD 136010$^1$ & galactic star & K0 II   \\ 
HD 136406$^1$ & galactic star & K0 III  \\
NGC 6356$^1$ & globular cluster & G3   \\
NGC 6809$^1$ & globular cluster & F4   \\
NGC 6981$^1$ & globular cluster & F4   \\ 
NGC 1851$^1$ & globular cluster & F4    \\
HD 223647$^2$ & galactic star & G7 III \\
47 Tucanae$^2$ & globular cluster & G4 \\
NGC 2298$^2$ & globular cluster & F5 \\
HD 35410$^2$ & galactic star & G9 III-IV \\
\hline
\end{tabular}
\end{table}

\section{Radial Velocities}

\begin{figure}
\epsfysize 3.0truein
\hfil{\epsffile{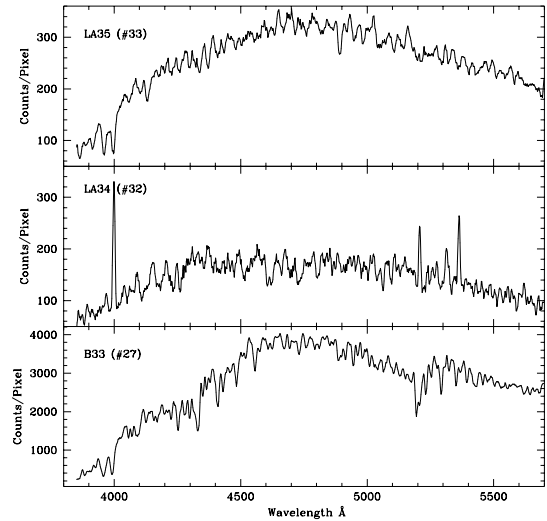}}\hfil
\caption{Spectra of globular cluster candidates around NGC~253. From top to 
bottom they are identified as: a globular cluster, $v_{\rm h}$ = 228 
$\pm$ 85 kms$^{-1}$, a background galaxy, $v_{\rm h}$ $=$ 19830 $\pm$ 
7 kms$^{-1}$
and a foreground galactic star, $v_{\rm h}$ = 5 $\pm$ 7 kms$^{-1}$. 
The galaxy shows emission lines of [O\,{\sc ii}], [O\,{\sc iii}] and H$\beta$. 
The spectra have been smoothed with a 3--pixel boxcar filter.}
\label{fig:spectra}
\end{figure}

Radial velocities were determined from the spectra through two methods. 
For obvious emission--line objects, identified lines were fitted with a 
gaussian
and the position of the peak measured. The final velocity is the
mean of these measurements for different lines, with the uncertainty being 
the rms between measurements.
For spectra with no obvious emission lines, radial velocities were determined 
by the cross--correlation of object spectrum against template \cite{tonry} 
with the task \textsc{fxcor} in \textsc{iraf}.
Emission lines more than 4~$\sigma$ from the psuedo--continuum 
(determined by fitting a polynomial to the spectra) where either 
interpolated across by hand or removed with the \textsc{lineclean} 
task in \textsc{iraf}.
Fig.~\ref{fig:spectra} shows three extracted spectra of globular cluster 
candidates in NGC~253.

It was required that the cross--correlations have normalised peak heights 
$>$ 0.1; below this threshold it was found that returned velocities were 
unreliable and should therefore be discarded. Furthermore it was specified 
that each spectrum should have at least two reliable cross--correlations 
against
different templates. The final velocities are taken to be the mean velocity
weighted by the cross--correlation peak height of each template, and have been 
corrected to heliocentric values. The uncertainties given are the rms between 
velocities derived from these different templates.
Tables \ref{tab:n253} and \ref{tab:n55} show the final velocities obtained 
from the candidate cluster spectra for the two galaxies.
There is no real information as to the nature of external uncertainties; 
none of these objects have been previously observed spectroscopically and 
since each field was observed solely in one setup, there are no overlaps 
between objects for comparison. 

\begin{table*}
\centering
\begin{minipage}{140mm}
\caption{Globular cluster candidates in NGC~253. Columns are: ID 
(fibre \#), heliocentric velocity, velocity error, right ascension, 
declination, other designation, $B$ magnitude and classification. }
\label{tab:n253}
\begin{tabular}{lllcclll}

\hline
ID & $v_{\rm h}$  &  $v_{\rm err}$ & $\alpha$(1950) & $\delta$(1950) & 
Alternative ID\footnote {LA $\equiv$ Liller \& Alcaino (1983b). 
LAb \&  LAc are 'blue' candidates from their sample. 
LAA, LAB \& LAC are labelled as 'bright' objects. 
B $\equiv$ Blecha (1986). } & $B$\footnote 
{magnitudes from  Liller \& Alcaino (1983b) and Blecha (1986)} & Classification\\
\hline

1 & 62081 & 21 & 00 45 55.42 & -25 14 19.1 & LA33 & 18.9 &	galaxy		\\
2 & 33731 & 79 & 00 45 17.95 & -25 24 59.1 & B3 & 20.1     &	galaxy          \\
4 & 23061 & 117 & 00 45 19.30 & -25 24 02.1 & B2 & 21.2   &	galaxy          \\
5 & 29601 & 30 & 00 45 19.29 & -25 19 11.2 & LA38 & 18.8 & 	galaxy		\\ 
7 & 212 & 74 & 00 45 35.09 & -25 26 58.8 & B1 & 21.1     &	cluster         \\
8 & 235 & 28 & 00 45 45.79 & -25 22 18.2 & LA26 & 20.0    &	cluster         \\
9 & 90390 & 12 & 00 46 02.22 & -25 15 54.7 & LAc & 19.4     &	galaxy          \\
10 & 44826 & 21 & 00 45 46.53 & -25 13 02.9 & LA41 & 19.6 & 	galaxy		\\
11 & 75687  & 21 & 00 45 00.52 & -25 22 02.2 & B5 & 20.8 & 	galaxy		\\ 
12 & 266 & 12 & 00 44 50.23 & -25 23 24.9 & LA40 & 19.3  &	cluster         \\
13 & 6552 & 70 & 00 44 39.18 & -25 29 28.9 & B7 & 20.1  &	galaxy          \\
14 & 64590 & 18 & 00 44 39.52 & -25 23 43.2 & B8 & 20.4  &	galaxy          \\
17 & 49540 & 89 & 00 45 33.02 & -25 15 14.1 & LA43 & 19.6  &	galaxy          \\
18 & 40825 & 96 & 00 44 58.63 & -25 29 35.0 & B6 & 21.4  &	galaxy          \\
19 & 10230 & 84 & 00 44 44.08 & -25 31 32.6 & B10 & 21.6  &	galaxy          \\
20 & 17940 & 46 & 00 45 54.02 & -25 37 01.9 & B30 & 19.1 & 	galaxy 		\\
21 & 28484 & 32 & 00 45 58.44 & -25 31 42.2 & B31 & 19.6 & 	galaxy		\\
22 & 16857 & 52 & 00 46 11.56 & -25 33 27.1 & LAA & 16.0  &	galaxy          \\
23 & 72749 & 21 & 00 46 22.16 & -25 32 31.8 & LA12 & 20.0 &	galaxy		\\
24 & 55775 & 117 & 00 45 56.23 & -25 32 31.3 & LA20 & 19.5 &	galaxy		\\
25 & 11 & 10 & 00 46 02.64 & -25 33 47.8 & LA17 & 18.4  &	star           \\
26 & 64353 & 27 & 00 45 38.92 & -25 31 33.8 & B32 & 20.9 &	galaxy		\\ 
27 & 5 & 7 & 00 45 49.27 & -25 31 00.7 & B33 & 16.5  &		star           \\
28 & 228 & 85 & 00 46 17.91 & -25 34 48.4 & LA11 & 20.1  &	cluster         \\
29 & 28629 & 50 & 00 45 57.56 & -25 35 47.1 & LAB & 16.8 & 	galaxy   	\\
30 & 44961 & 90 & 00 43 51.17 & -25 30 12.4 & LA51 & 18.1  &	galaxy          \\
31 & 41919 & 102 & 00 44 03.39 & -25 25 34.5 & LA52 & 19.2 &	galaxy		\\
32 & 19830 & 7 & 00 43 50.18 & -25 36 11.2 & LA34 & 19.2  &	galaxy          \\
33 & 230 & 14 & 00 43 47.77 & -25 36 29.0 & LA35 & 17.6  &	cluster         \\
35 & 33206 & 167 & 00 43 51.01 & -25 38 50.1 & LA29 & 19.6  &	galaxy          \\
36 & 352 & 26 & 00 44 18.06 & -25 34 05.8 & B15 & 21.0  &	cluster         \\
38 & 70834 & 30 & 00 44 14.58 & -25 20 39.9 & LA56 & 20.0 &	galaxy		\\
39 & 339 & 56 & 00 44 13.07 & -25 20 51.9 & LA57 & 20.3  &	cluster         \\
40 & 401 & 117 & 00 45 05.84 & -25 39 53.7 & LA24 & 19.4  &	cluster         \\
41 & 218 & 22 & 00 45 20.65 & -25 34 03.7 & B29 & 22.3  &	cluster         \\
42 & 313 & 85 & 00 45 05.14 & -25 37 56.2 & B24 & 21.0  &	cluster         \\
43 & 40796 & 187 & 00 45 52.57 & -25 42 13.9 & LA8 & 19.3  &	galaxy          \\
44 & 447 & 102 & 00 46 04.90 & -25 45 18.7 & LA3 & 19.7  &	cluster         \\
45 & 57852 & 59 & 00 45 45.80 & -25 39 51.5 & B27 & 19.5  &	galaxy          \\
46 & 89151 & 24 & 00 45 20.39 & -25 34 59.1 & B28 & 21.5  &	galaxy          \\
47 & 38485 & 300 & 00 45 24.26 & -25 36 59.2 & B25 & 21.2 & 	galaxy		\\
50 & 28236 & 4 & 00 44 44.24 & -25 40 59.8 & B22 & 21.1	&	galaxy		\\ 
51 & 10233 & 91 & 00 44 38.69 & -25 34 57.5 & LAb & 19.4  &	galaxy          \\
52 & 20863 & 82 & 00 44 49.22 & -25 32 06.3 & B11 & 19.6  &	galaxy          \\
54 & 22927 & 27 & 00 44 34.79 & -25 40 47.3 & B20 & 18.8  &	galaxy          \\
55 & 9865 & 364 & 00 44 19.47 & -25 40 00.8 & B19 & 20.7  &	galaxy          \\
56 & 39522 & 102 & 00 44 38.88 & -25 34 26.6 & B16 & 21.6  &	galaxy          \\
57 & 361 & 45 & 00 44 55.95 & -25 32 03.2 & B12 & 20.6  &	cluster         \\
58 & 261 & 52 & 00 44 55.64 & -25 32 44.6 & B13 & 21.4  &	cluster         \\
60 & 21522 & 76 & 00 43 48.08 & -25 38 32.7 & LAC & 16.1  &	galaxy          \\
61 & 302 & 22 & 00 44 55.98 & -25 33 03.1 & B14 & 20.2  &	cluster         \\
62 & 71064 & 15 & 00 44 34.94 & -25 36 03.6 & B18 & 21.2 &	galaxy		\\ 
63 & 21020 & 7 & 00 44 32.99 & -25 43 48.5 & B21 & 20.9  &	galaxy          \\
64 & 406 & 10 & 00 44 47.76 & -25 38 49.3 & B23 & 20.3  &	galaxy          \\

\hline
\end{tabular}
\end{minipage}
\end{table*}

 \begin{table*}
\centering
\begin{minipage}{140mm}
\caption{Globular cluster candidates in NGC~55. Columns are: ID 
(fibre \#), heliocentric velocity, velocity error, right ascension, 
declination, other designation, $B$ magnitude and classification.} 
\label{tab:n55}
\begin{tabular}{lllcclll}

\hline
ID & $v_{\rm h}$ &  $v_{\rm err}$ & $\alpha$(1950) & $\delta$(1950) & 
Alternative ID\footnote{LA $\equiv$ Liller \& Alcaino (1983a). LAb \& LAd 
are labelled as 'blue' objects in their paper. 
LAA, LAB \& LAC are labelled 'bright' objects.
Objects here denoted n,q,k,o,h,g,t,a \& f were considered by Liller \& 
Alcaino as being too blue for classical globular clusters, 
but were included for completeness. 
B $\equiv$ Blecha (1986).} & $B$\footnote {magnitudes from Liller \& 
Alcaino (1983a) and Blecha (1986)} & Classification\\
\hline

1 & 66700 & 108 & 00 12 50.86 & -39 17 25.9 & LA61 & 19.0 &	galaxy\\
2 & 35229 & 101 & 00 12 43.09 & -39 16 42.3 & n & -- &		galaxy\\
3 & 68 & 4 & 00 12 40.60 & -39 11 45.0 & q & -- & 		star\\
5 & 48 & 49 & 00 12 26.54 & -39 22 12.1 & k & -- & 		star\\
7 & 32936 & 89 & 00 11 06.35 & -39 39 43.8 & LA5 & 19.8 & 	galaxy\\
8 & -58 & 80 & 00 12 43.35 & -39 14 59.8 & o & -- & 		star\\
11 & -- & -- &  00 11 19.98 & -39 24 55.2 & LA46 & 19.6 & 	QSO \\
12 & 49388 & 136 & 00 11 22.17 & -39 24 23.3 & LA48 & 19.6 & 	galaxy \\
13 & 34933 & 84 & 00 11 05.77 & -39 20 49.3 & h & -- & 		galaxy\\
15 & 26170 & 194 & 00 11 02.95 & -39 16 20.8 & LA58 & 20.7 & 	galaxy\\
16 & 38076 & 123 & 00 11 09.12 & -39 23 50.0 & LA47 & 18.2 & 	galaxy\\
17 & 19941 & 194 & 00 12 10.22 & -39 26 17.8 & g & -- & 	galaxy\\
21 & 52918 & 18 & 00 11 08.98 & -39 34 11.9 & LA27 & 20.3 & 	galaxy\\
23 & 96708 & 24 & 00 11 51.76 & -39 32 44.5 & LA36 & 20.6 &	galaxy \\
24 & 52935 & 24 & 00 11 12.27 & -39 34 55.6 & LA25 & 20.1 & 	galaxy \\
25 & 17282 & 109 & 00 11 18.63 & -39 36 06.9 & LA20 & 19.7 & 	galaxy\\
26 & 88804 & 24 & 00 11 13.40 & -39 35 32.8 & LA21 & 20.2 & 	galaxy\\ 
27 & -60 & 72 & 00 12 03.76 & -39 29 43.2 & LA45 & 19.8 & 	star\\
29 & 35624 & 21 & 00 11 40.46 & -39 32 56.6 & LA32 & 18.9 & 	galaxy\\
30 & 30855 & 94 & 00 12 28.05 & -39 42 33.8 & LA11 & 18.3 & 	galaxy\\
31 & 26016 & 179 & 00 12 25.96 & -39 42 50.8 & LA9 & 20.5 & 	galaxy\\
32 & 35060 & 14 & 00 12 38.50 & -39 39 20.7 & LA30 & 19.3 & 	galaxy\\ 
33 & 47568 & 90 & 00 12 37.40 & -39 40 28.2 & LA26 & 19.5 & 	galaxy\\
35 & 48265 & 18 & 00 12 09.95 & -39 39 53.0 & LA18 & 19.7 & 	galaxy \\
36 & 71400 & 200 & 00 12 21.16 & -39 42 38.3 & LA8 & 20.2 & 	galaxy \\
38 & 49814 & 108 & 00 12 19.71 & -39 40 39.9 & LA16 & 18.6 & 	galaxy\\
39 & 30811 & 70 & 00 12 55.13 & -39 41 44.8 & t & -- & 		galaxy\\
41 & 75000 & 144 & 00 12 57.40 & -39 42 10.0 & LA22 & 19.6 & 	galaxy \\
42 & 74215 & 24 & 00 12 46.90 & -39 44 50.0 & LAd & -- & 	galaxy\\
43 & 81546 & 103 & 00 11 55.57 & -39 43 20.9 & LA3 & 20.2 & 	galaxy\\
44 & 32767 & 24 & 00 11 14.67 & -39 40 35.8 & LAb & -- & 	galaxy \\
46 & 60311 & 305 & 00 11 54.10 & -39 40 35.8 & LA10 & 20.2 & 	galaxy\\
47 & 30857 & 74 & 00 13 03.05 & -39 43 37.4 & LA15 & 16.9 & 	galaxy\\
48 & 30868 & 62 & 00 13 00.49 & -39 43 34.3 & LA14 & 17.9 & 	galaxy\\
49 & 60695 & 18 & 00 11 50.07 & -39 46 19.8 & a & -- & 		galaxy\\
51 & 5089 & 122 & 00 12 52.95 & -39 33 55.0 & LA42 & 20.3 & 	galaxy\\
52 & 52710 & 177 & 00 12 46.37 & -39 35 20.7 & LA39 & 19.2 & 	galaxy\\
53 & 246 & 58 & 00 13 17.91 & -39 35 19.7 & LA43 & 17.1 & 	cluster\\
54 & 37210 & 118 & 00 12 57.89 & -39 36 43.5 & LA37 & 20.4 & 	galaxy \\
55 & 75407 & 27 & 00 12 24.29 & -39 35 37.7 & LA34 & 19.8 & 	galaxy\\
56 & 92448 & 24 & 00 13 35.81 & -39 25 23.7 & LA59 & 20.3 & 	galaxy\\
57 & -142 & 70 & 00 12 28.78 & -39 37 46.9 & LA31 & 20.3 & 	star\\
58 & -111 & 23 & 00 13 23.85 & -39 29 50.7 & LA51 & 18.8 & 	star\\
59 & 75971 & 104 & 00 13 29.40 & -39 36 33.2 & LA41 & 20.3 & 	galaxy\\
60 & 31305 & 18 & 00 13 12.09 & -39 37 15.9 & f & -- & 		galaxy\\
61 & 8763 & 39 & 00 11 55.37 & -39 40 07.5 & LAB & 16.4 & 	galaxy\\
62 & 32793 & 111 & 00 11 16.82 & -39 39 21.0 & LAA & 16.7 & 	galaxy\\
63 & -29 & 56 & 00 13 02.95 & -39 23 51.2 & LAC & 16.1 & 	star\\
64 & 56585 & 12 & 00 12 44.39 & -39 35 03.3 & LA40 & 20.6 & 	galaxy\\

\hline
\end{tabular}
\end{minipage}
\end{table*}

\section{Identification of Globular Clusters}

All our globular cluster candidates have been classified as extended objects
on the basis of visual inspection of photographic plates (Liller \& Alcaino 
1983a,b) or profile analysis (Blecha 1986).
The main source of contamination in the samples should therefore be from 
background galaxies.
The systemic velocity of the Sculptor group is low, 
($V_{\rm N55}=129$~kms$^{-1}$; $V_{\rm N253}=245$~kms$^{-1}$), 
so we take a velocity cut at $v_{\rm h} = $ 1000~kms$^{-1}$, 
with all objects above this threshold being assumed to be  background 
galaxies. 
A large number of the spectra exhibited emission lines of  
\hbox{[O\,{\sc ii}]}, \hbox{[O\,{\sc iii}]}, H${ \beta}$ or H$\gamma$ and  
except for candidate \#64 in NGC~253 (see below), these objects 
all had $v_{\rm h} > $ 1000~kms$^{-1}$.
Removing background galaxies in such a manner should effectively 
leave a sample consisting of objects which are either residual 
contaminating foreground Galactic stars, or globular clusters within the 
Sculptor group. 
For old clusters in the Milky Way, Armandroff (1989) gives a velocity 
dispersion 
of $\sigma$ =  100 kms$^{-1}$. Assuming that mass scales with  $\sigma^{2}$ 
and the mass--to--light ratios of the Milky Way and the two Sculptor 
spirals are comparable, then the expected velocity dispersion for the 
halo clusters in NGC~55 and NGC~253 will be $\sim$ 70 kms$^{-1}$. 
The velocity ranges for possible clusters ( $\pm$ 3~$\sigma$ from the 
mean velocity) are then taken to be 
-80 $ \le V_{\rm N55} \le $ 340~kms$^{-1}$ and 35 $ \le  V_{\rm N253} \le $ 
455~kms$^{-1}$ for NGC~55 and NGC~253 respectively.
Since at some level there will be an overlap between high velocity 
foreground stars and globular clusters in the Sculptor group, distinguishing
between these two cases relies on identifying true clusters as 
marginally extended objects.

Fifteen NGC~253 cluster candidates fall within the velocity range expected 
for globular clusters, and appear marginally resolved on images from the 
Digitized Sky Survey (see Lasker \& Mclean 1994). 
Object \#64 has $v_{\rm h} = $ 404 $\pm$ 10~kms$^{-1}$, but shows emission 
lines of \hbox{[O\,{\sc iii}]}, H${ \beta}$ and H$\gamma$ and has 
therefore been excluded from further consideration. 
Two of the candidates, \#25 and \#27, have radial velocities which 
fall just short of the velocity cut. The appearance of their images and 
their COSMOS image parameters (see \S 5) point towards a Galactic origin; 
on this basis they have been identified as foreground stars.
Objects \#40 and \#44 both have large radial velocities in comparison 
to the mean systemic velocity of NGC~253, with $v_{\rm h} = $ 401 $\pm$ 
117~kms$^{-1}$ 
and $v_{\rm h} = $ 447 $\pm$ 102~kms$^{-1}$ respectively. However, the large 
errors reflect the significant degree of scatter between templates where 
the cross--correlations were close to the normalised peak height cut--off at 
0.1 and so they have been left in the sample. Table \ref{tab:clusters} lists 
those objects in NGC~253 that are identified as globular clusters.
The sample has mean velocity $\bar{v_{\rm h}} = $ 297 kms$^{-1}$ with velocity 
dispersion, $\sigma =$ 74 kms$^{-1}$. Omitting objects \#40 and \#44 yields  
$\bar{v_{\rm h}} = $ 276 kms$^{-1}$ and $\sigma =$ 55 kms$^{-1}$, this is 
entirely consistent with values expected for the NGC~253 globular 
cluster system. Twelve of the clusters fall into the 'classical' 
colour region for globular clusters, with 0.5 $< B-V <$ 1.25.
Cluster \#7 (B1) is very blue, with $B-V =$ 0.19. Its spectrum shows 
strong Balmer absorption lines, and may be analogous to the blue clusters 
seen in the Magellanic Clouds (e.g. Bica, Dottori, \& Pastoriza 1986). 
Cluster \#41 has $B-V =$ 1.8, and is probably highly reddened due to its 
proximity to the disk of NGC~253 (see Blecha 1986).  

In the NGC~55 sample, there are six objects that are not galaxies 
and fall within the velocity cut for globular clusters. However, all 
but one of these appear effectively point--like  and lie on the stellar 
locus of the COSMOS plate scans (see \S 5). Object \#53 has $v_{\rm h} = 
$ 246 $\pm$ 58 kms$^{-1}$ and lies within 2~$\sigma$ of the galaxies' 
systemic velocity. Images of this object show some elongation, 
but an isophotal plot indicates a round, marginally extended source 
blended with another object.\footnote{ In their catalogue, Liller \& Alcaino 
\shortcite{laa} indicate that object \#53 (LA43) is separated by 5$\farcs$9
from a fainter companion.} On the basis of this, it has been classified as a 
likely globular cluster. At $B = 17.1~(B-V = 0.76)$, cluster \#53 is bright 
but not unreasonably so. Adopting a distance modulus to NGC~55 of 
$(m-M)_{o}= $ 26.5 gives the cluster $M_{V}$ $=$ -10.2, the same as the  
most luminous Galactic globular cluster $\omega$ Cen \cite{harris96}.

\begin{figure}
\epsfysize 3.0truein
\hfil{\epsffile{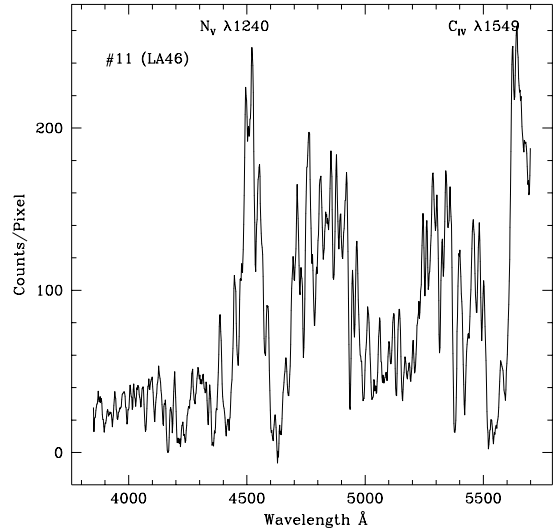}}\hfil
\caption{ Object \#11 in the NGC~55 cluster sample, identified as a broad 
absorption line (BAL) QSO at $z \sim$ 2.7. Emission in N\,{\sc v} and 
C\,{\sc iv} is indicated, with their corresponding UV rest--frame wavelengths.}
\label{fig:qso}
\end{figure}

Object \#11 has been identified as a broad absorption line (BAL) QSO with
redshift $z \sim $2.7. Its spectrum is shown in Fig.~\ref{fig:qso}. 
Emission in \hbox{N\,{\sc v}} and \hbox{C\,{\sc iv}} is shown, 
though absorption shortward of \hbox{N\,{\sc v}} is so strong that no 
Ly$\alpha$ emission is observed. This type of spectrum is occasionally seen 
in 'peculiar' BAL QSOs (e.g. Korista et al. 1995).
 
\begin{table}
\centering
\caption{Objects in the NGC~253 sample identified as 
globular clusters. From left to right the columns give: ID, radial velocity, 
velocity error, other identification, $B$ magnitude, $B-V$ colour and COSMOS
ellipticity.}
\label{tab:clusters}
\begin{tabular}{lllllll}

\hline
ID & $v_{\rm h}$ &  $v_{\rm err}$ & Other ID & $B$ & $B-V$ & $\epsilon$\\
\hline

7 & 212 & 74 &  B1 & 21.1 & 0.19 & -- \\
8 & 235 & 28 &  LA26 & 20.0 & 0.68 & 0.28 \\
12 & 266 & 12 & LA40 & 19.3 & 0.89 & 0.07 \\       
28 & 228 & 85 & LA11 & 20.1 & 0.70 & 0.12 \\
33 & 230 & 14 & LA35 & 17.6 & 0.77 & -- \\
36 & 352 & 26 & B15 & 21.0 & 0.94 & 0.23 \\
39 & 339 & 56 & LA57 & 20.3 & 0.85 & 0.08\\
40 & 401 & 117 & LA24 & 19.4 & 0.97 & -- \\
41 & 218 & 22 & B29 & 22.3 & 1.80 & 0.22 \\
42 & 313 & 85 & B24 & 21.0 & 0.86 & 0.22 \\
44 & 447 & 102 & LA3 & 19.7 & 0.94 & -- \\
57 & 361 & 45 & B12 & 20.6 & 0.85 & 0.19 \\
58 & 261 & 52 & B13 & 21.4 & 0.74 & 0.10 \\
61 & 302 & 22 & B14 & 20.2 & 0.54 & -- \\

\hline
\end{tabular}
\end{table}

\begin{figure}
\epsfysize 3.0truein
\hfil{\epsffile{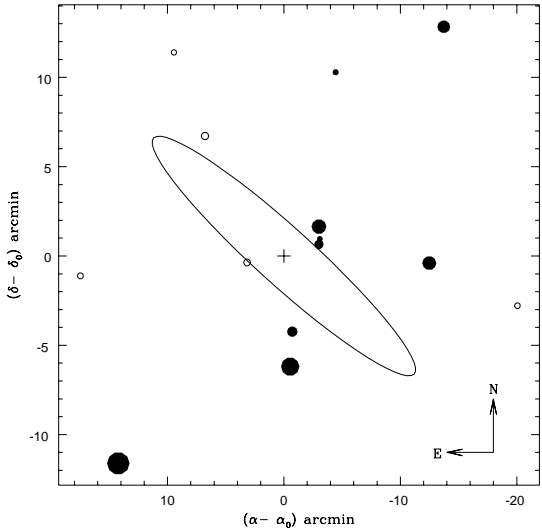}}\hfil
\caption{Distribution of globular clusters about NGC~253. Symbol size 
represents the magnitude of the globular cluster radial velocity with 
respect to the galaxy rest frame (systemic velocity $V_{\rm N253} =$ 245 
kms$^{-1}$). Open circles are approaching clusters, filled circles 
indicate receding clusters in this scheme. The centre of the coordinate 
system is that of NGC~253, $\alpha(1950) = 00\degr 45\arcmin 08\arcsec$, 
$\delta(1950) = -25\degr 33\arcmin 42\arcsec$.
The ellipse indicates the orientation of the galaxy on the sky, 
with a position angle of the major axis of 50$^\circ$ (Pence 1980).}
\label{fig:distribution}
\end{figure}

Fig.~\ref{fig:distribution} shows the spatial distribution of the 14 objects 
identified as globular clusters in NGC~253. The field is aproximately 
40~arcminutes on a side.
The distribution of cluster velocities is interesting; the clusters 
predominantly recede with respect to the galaxy rest--frame in the SW part 
of the galaxy and approach in the NE. This is consistent 
with the direction of rotation of the galaxy, as measured from H$\alpha$ 
 rotation curves \cite{pence}. However, due to the small numbers of 
clusters the level of rotation is not statistically significant.

\section{Definition of a New Sample of Globular Clusters}

\subsection{COSMOS plate scans}

The cluster samples selected by Liller \& Alcaino (1983a,b) are based on 
visual inspection of photographic plates. Clearly such selection is prone to
 subjective 
errors and saturation effects. A more quantitative approach was taken by 
Blecha (1986) using electronographic plates, but as can be seen from 
Table \ref{tab:n253}, even this sample is contaminated by both 
foreground stars and background galaxies. In this section we explore the use 
of image parameters measured by the COSMOS measuring machine 
(Beard, MacGillivray \& Thanisch 1990) to identify new samples of 
globular clusters around NGC~253 and NGC~55.

AAT prime focus plates of the two Sculptor group galaxies were raster
scanned with the COSMOS facility using the mapping mode, with a step size of 
16 $\mu$m and a 16 $\mu$m spot size. The image area threshold was set 
to 10 pixels (plate scale 15.3 arcsec mm$^{-1}$), with all pixels above 
this threshold being grouped into discrete objects. 
Details of the plates are listed in Table \ref{tab:plates}.  

\begin{table}
\centering
\caption{Prime focus plates scanned using COSMOS}
\label{tab:plates}
\begin{tabular}{lllccll}

\hline
Plate & Object & Date & Exp. & Emulsion & Filter \\
 & & &  (min) & & & \\
\hline

1739 & NGC~253 & 20--10--79& 80 & IIIa--J & GG385 \\
1740 & NGC~55 & 20--10--79 & 90 & IIIa--J & GG385 \\

\hline
\end{tabular}
\end{table}

\begin{figure}
\epsfysize 3.0truein
\hfil{\epsffile{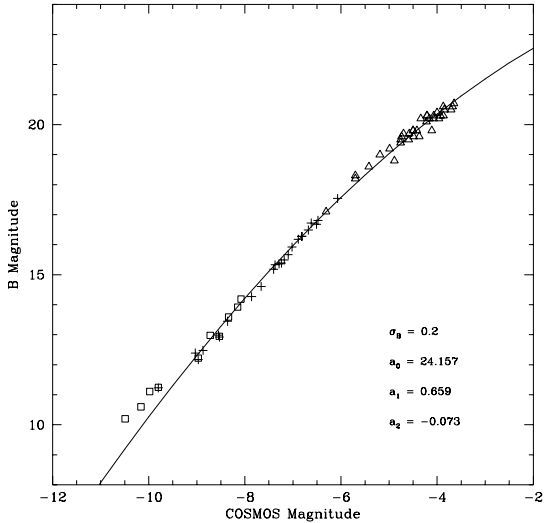}}\hfil
\caption{NGC~55: Correlation between COSMOS instrumental magnitudes and 
photometry of Liller \& Alcaino (1983a), Alcaino \& Liller (1984) and 
Hanes \& Grieve (1982). 
Alcaino \& Liller photometric stars are shown as crosses, Hanes \& Grieve 
stars are open squares. Cluster candidates are shown as open triangles. 
A least--squares fit to the data gives a formal error in $B$ of $\sigma =$ 
0.2 mag. }
\label{fig:cal55}
\end{figure}

\begin{figure}
\epsfysize 3.0truein
\hfil{\epsffile{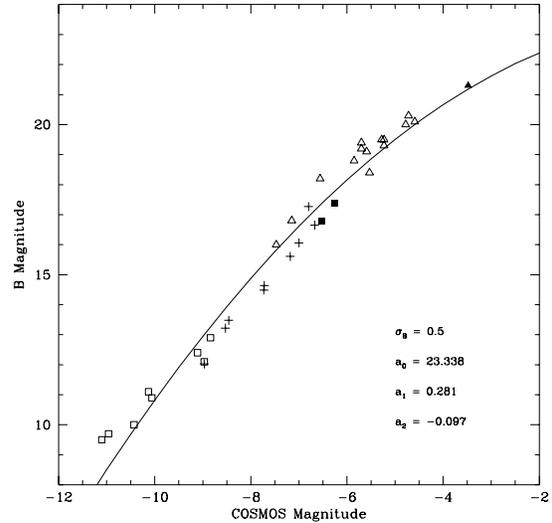}}\hfil
\caption{NGC~253: Correlation between COSMOS instrumental magnitudes 
and photometry of Liller \& Alcaino (1983b) (open triangles), 
Alcaino \& Liller (1984) (crosses) and Hanes \& Grieve (1982) (open squares). 
Also plotted as a filled triangle is a faint star 
($m_{B} =$ 21.3) from the NLTT catalogue (Luyten 1980) and two
 additional stars 
from the SIMBAD database (filled squares). 
A least--squares fit to the data gives a formal error in $B$ of $\sigma =$ 
0.5 mag.}
\label{fig:cal253}
\end{figure}

The COSMOS image analysis software \cite{stobie} was then run on each 
digitized frame to provide a list of objects with information on position, 
magnitude, orientation, axial ratio (major and minor axis lengths) and area.
The COSMOS instrumental magnitudes were calibrated using the photoelectric 
sequences of Hanes \& Grieve (1982), and Alcaino \& Liller (1984).\footnote 
{It should be noted for future reference that the star designated 
as V on their finding chart 38 for NGC~253 is tabulated as P in the 
paper.} Alcaino \& Liller give $V$ magnitudes and colours for 24 stars 
down to $V$ = 16.95 ($B-V$ = 0.59) in the field of NGC~55, and for 19 
stars to $V$ = 16.53 ($B-V$ = 0.74) in the vicinity of NGC~253. 
Those from Hanes \& Grieve are somewhat brighter, in the range 
9 $\leq V \leq$ 14 for the two galaxies. 
There are a total of ten overlaps between their photometric sequences, 
with six in the NGC~253 field and four in the NGC~55 field. 
Both give good agreement, with a mean $B$ magnitude offset, $\Delta_{B} = 
0.025$~mag, $\sigma  = 0.06$ and $\Delta_{B} 
= 0.045$~mag, $\sigma  = 0.03$ for NGC~253 and NGC~55 respectively.
For calibration at the faint end, the cluster candidates reached 
down to $B \sim 21$, and a faint star  ($m_{B} = 21.3$) from the New
Luyten Two--Tenths catalogue (NLTT) \cite{luyten} was also used for 
the calibration of plate J1739. 
\footnote{ There is significant disagreement between Blecha's and 
Liller \& Alcaino's photometry. In some instances the data differs by a 
full magnitude for the same object. This is seen in the COSMOS data 
where Blecha's candidates are systematically fainter -- see Blecha 
(1986) for more details. For internal consistency, only 
Liller \& Alcaino's photometry was used for the calibration of the J1739 
plate.}
Fig.~\ref{fig:cal55} shows the best--fitting calibration curve for the 
COSMOS instrumental magnitudes for NGC~55. 
A least--squares fit to the data yields an rms of 0.2~mag. 
The photometric calibration for NGC~253 shown in Fig.~\ref{fig:cal253} 
shows greater scatter, with $\sigma = 0.5$~mag. 
 The origin of this larger scatter in the NGC~253 plate is unknown, 
but is seen over the entire magnitude range.

An astrometric solution for the COSMOS scans was obtained  
so as to locate spectroscopically identified objects from this work 
within the COSMOS datasets. Typical astrometry plate residuals were 
of order $\sim$ 0.3 arcseconds.  

\subsection{A New Cluster Sample}

Globular clusters should appear as round, marginally extended objects on the 
photographic plates, whereas stars should be effectively point--like.

The ellipticity for each COSMOS object is derived from its measured 
image moments, where ellipticity is defined as $\epsilon = 1 - b/a$ 
($b$ and $a$ are the measured semi--minor and semi--major axes respectively).
For both plates an ellipticity limit was set at approximately three times 
the rms stellar ellipticity, above which threshold objects were classified as 
being either truly elliptical or a blend of two or more objects. These were 
consequently excluded from the new cluster candidate sample.
Object ellipticity showed only a weak dependence on magnitude, in the 
sense that
 at fainter magnitudes the numbers of elliptical objects \emph{marginally} 
increased. 
This was not a strong trend and fainter globular clusters should not be 
discriminated against in this scheme (the limits which define elliptical 
images increase towards fainter magnitudes.)

\begin{figure}
\epsfysize 3.0truein
\hfil{\epsffile{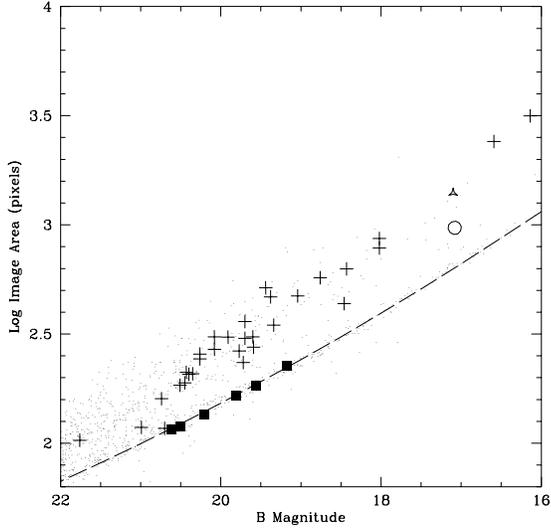}}\hfil
\caption{Log image area versus magnitude for objects in NGC~55. COSMOS 
objects are plotted as points, confirmed galaxies are crosses and stars 
are represented by filled squares. 
The open triangle indicates cluster \#1 of Da Costa \& Graham (1982) and 
the position of the globular cluster identified in this work is indicated 
by an open circle. The dashed line shows the fit applied to the stellar 
sequence.}
\label{fig:n55log}
\end{figure}

\begin{figure}
\epsfysize 3.0truein
\hfil{\epsffile{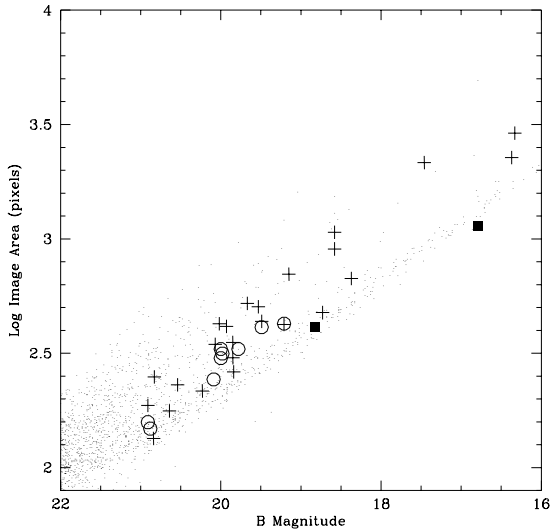}}\hfil
\caption{Log image area versus magnitude for objects in NGC~253. COSMOS 
objects are plotted as points, confirmed galaxies are crosses, stars 
are represented by filled squares and open circles indicate confirmed 
globular clusters.}
\label{fig:n253log}
\end{figure}

A powerful diagnostic to test for the extended nature of the candidates 
is the measured COSMOS image area as a function of magnitude. 
Figs. \ref{fig:n55log} and \ref{fig:n253log} show the logarithm of the 
image area (in pixels) against magnitude for COSMOS objects in the magnitude 
range $16 \leq B \leq 22$. The sharply defined sequence is the stellar 
locus arising from foreground stars. 
Extended objects show an excess of area for their magnitude, and are raised 
from this sequence to some degree. 
Spectroscopically identified objects from this work which were located in 
the COSMOS data are shown. 
Da Costa \& Graham (1982) identify three bright star clusters with $V \sim$ 
17 near the centre of NGC~55, one of which was returned in the COSMOS data 
and is indicated in Fig.~\ref{fig:n55log}. The two remaining clusters, 
along with a number of cluster candidates failed to be found. 
These objects are all seen in projection close to the disks of the two 
galaxies, 
and were presumably lost in the bright local background. Also, it should 
be noted that whilst the spectroscopic sample went out to radii 
comparable with that of the plate scans, NGC~253 was slightly offset from the 
centre of the COSMOS field, and consequently some candidates fell outside 
the scanning limits.

\begin{figure}
\epsfysize 3.0truein
\hfil{\epsffile{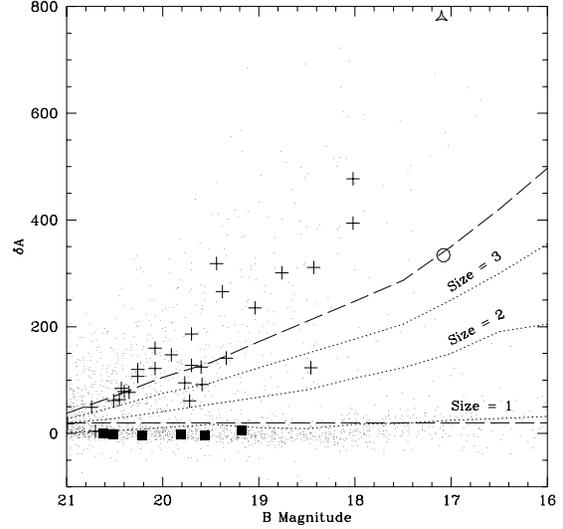}}\hfil
\caption{Area residuals of round candidates in NGC~55. The residual 
$\delta A = $0 corresponds to the stellar locus. Confirmed galaxies are 
shown as crosses, stars are indicated by filled squares. The probable 
globular cluster is shown as an open circle. Dotted lines are the cluster 
models of Harris et al. (1984). The long dashed lines 
indicate the parameter space containing new globular cluster candidates.}
\label{fig:resid55}
\end{figure}

\begin{figure}
\epsfysize 3.0truein
\hfil{\epsffile{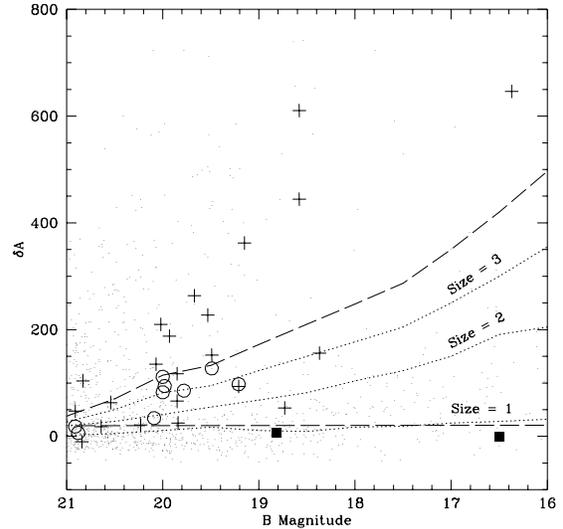}}\hfil
\caption{Residuals in area for objects in the NGC~253 field. Symbols are as
for previous figure. }
\label{fig:resid253}
\end{figure}

In order to quantify the excess in area shown by genuinely extended objects, 
and to differentiate between object types (stars, galaxies and globular 
clusters) a line was fit to the stellar sequence and a residual area, 
$\delta A$, for each object calculated.
Figs.~\ref{fig:resid55} and~\ref{fig:resid253} show the results of this 
exercise, where $\delta A = $ 0 corresponds to the  stellar locus.
Several points are evident from these figures. The scatter in $\delta A$ is 
somewhat greater in NGC~253 than in NGC~55, with its the stellar locus 
being less well constrained (reflecting perhaps the larger photometry 
residuals -- see $\S4$). 
Those objects which lie on the stellar locus all have radial velocities 
consistent with foreground stars, and appear point--like on Sky Survey images.
Galaxies are located well above the stars, and follow a similar distribution 
for both fields, with their residual areas increasing strongly as a  
function of magnitude. Identified globular clusters largely inhabit a 
parameter space \emph{between} the stars and galaxies, though some overlap 
is apparent.

Overplotted are the model predictions of Harris et al. \shortcite{harris2}, 
scaled to the distance of the Sculptor group. These are produced 
by applying the COSMOS algorithm to a set of model cluster images with a 
range of magnitudes. The model clusters are all generated with King profiles 
\cite{king}. Those shown in Figs.~\ref{fig:resid55} and~\ref{fig:resid253} 
have $r_{t}/r_{c} = 30$, though models with other concentration parameters 
( $r_{t}/r_{c} = 6 - 100$) show similar characteristics. Model clusters shown 
with size = 1 possess core radii of 1.5 pc and tidal radii of 45 pc 
(0.15 and 4.6 arcsec respectively at 2 Mpc.) Sizes 2 and 3 are 
correspondingly twice and three times as large. Six of the spectroscopically 
identified globular clusters in NGC~253 lie below the size = 3 model line, 
with the remaining three lying marginally above this. The apparent
size of clusters on the sky will be sensitive to the assumed distance 
to the two galaxies, and as pointed out by Harris et al. \shortcite{harris2}, 
each plate exhibits slightly different behaviour due to saturation effects and
their individual isophotal thresholds (in fact they indicate that magnitudes 
can be in error by as much 0.6 mag for the most extended globular cluster
candidates.) The cluster of Da Costa \& Graham (1982) shown in Fig.
~\ref{fig:resid55} shows characteristics more like that of the galaxies 
than the other globular clusters, though DSS images of this object show an
extremely bright and crowded background which could possibly confuse 
the analysis software.\footnote{There is little chance of a misidentification 
by Da Costa \& Graham between their globular cluster and a background galaxy. 
They obtain a radial velocity for their cluster of $v_{\rm h} =  106 \pm 8$ 
kms$^{-1}$ and it is seen in projection against the stellar light of NGC~55.}

The regions of parameter space which should contain many globular clusters
are delimited in Figs.~\ref{fig:resid55} and~\ref{fig:resid253} by the 
long dashed lines. The lower limit (above the stellar locus) is set to 
be 3~$\sigma$ from the mean stellar area residual. The upper limit corresponds
 to approximately size = 4 model clusters and was chosen so as to include as 
many cluster candidates as possible, whilst minimizing contamination from 
background galaxies. 
The degree of crowding increases significantly at fainter magnitudes, 
therefore a magnitude cut at $m_{B}$ $\leq$ 20.5 has also been taken. 
As a final measure, 
images of each of the candidates have been visually examined for obvious 
stellar appearance or any structure indicative of a background
galaxy. Some 20 objects in total were identified as being obviously 
'non--cluster--like' and were removed from the sample. From the 
initial datasets, 91 cluster candidates have been identified in NGC~253 and 
84 in NGC~55. Table \ref{tab:candidates} lists the positions and 
approximate $B$ magnitudes of the new cluster candidate samples. 

\begin{table*}
\centering
\caption{New globular cluster candidates in NGC~253 and NGC~55}
\label{tab:candidates}
\begin{tabular}{llllllll}

\hline
NGC~253 & & & & NGC~55 & & & \\
ID & $\alpha$(1950) & $\delta$(1950) & $B$ & ID & $\alpha$(1950) & $\delta$(1950) & $B$ \\
\hline

1 & 0 44  7.68 & -25 39 16.1 & 18.37 & 1 & 0 13  3.46 & -39 38 55.7 & 18.69 \\  
2 & 0 46 44.19 & -25 29 40.4 & 18.46 & 2 & 0 12 57.83 & -39 26  3.6 & 18.70 \\  
3 & 0 46 59.32 & -25 30 24.6 & 18.47 & 3 & 0 14 24.19 & -39 15 25.7 & 18.72 \\  
4 & 0 43 54.34 & -25 9 47.0 & 18.53 & 4 & 0 10 47.66 & -39 37 41.6 & 18.72 \\  
5 & 0 44 26.32 & -25 34 44.3 & 18.54 & 5 & 0 15  0.86 & -39 35 45.1 & 18.86 \\  
6 & 0 47  1.08 & -25 39 17.8 & 18.60 & 6 & 0 11 29.17 & -39 19 43.2 & 18.87 \\  
7 & 0 46 54.30 & -25 36 15.5 & 18.67 & 7 & 0 11  1.67 & -39 53 41.0 & 18.87 \\  
8 & 0 44  8.25 & -25 22 19.4 & 18.68 & 8 & 0 13 48.19 & -39 14 29.4 & 18.91 \\  
9 & 0 44 47.79 & -25 32  9.6 & 18.76 & 9 & 0 12 42.25 & -39 1 49.5 & 18.93 \\   
10 & 0 46 13.33 & -25 28 37.7 & 18.80 & 10 & 0 11 42.33 & -39 25  8.1 & 19.03 \\  
11 & 0 45 17.67 & -25 9 38.2 & 18.83 & 11 & 0 12  8.84 & -39 25 59.1 & 19.08 \\ 
12 & 0 46 33.06 & -25 40 55.7 & 18.83 & 12 & 0 12  0.20 & -39 28 45.6 & 19.12 \\  
13 & 0 43 45.23 & -25 23  9.1 & 18.90 & 13 & 0 14  4.81 & -39 23 47.1 & 19.14 \\  
14 & 0 43 47.97 & -25 17 58.8 & 18.98 & 14 & 0 12 52.24 & -39 42 50.1 & 19.14 \\  
15 & 0 46 37.88 & -25 28 42.6 & 18.99 & 15 & 0 12 15.83 & -39 44 26.2 & 19.14 \\  
16 & 0 44 42.51 & -25 39 44.7 & 19.00 & 16 & 0 14 25.55 & -39 33 18.0 & 19.17 \\  
17 & 0 44 39.09 & -25 18 22.5 & 19.02 & 17 & 0 12 45.44 & -39 44 12.0 & 19.17 \\  
18 & 0 45 29.49 & -25 21 44.1 & 19.14 & 18 & 0 12 47.47 & -39 10 42.2 & 19.19 \\  
19 & 0 46 39.13 & -25 22 34.6 & 19.20 & 19 & 0 13  7.21 & -39 57 19.7 & 19.19 \\  
20 & 0 46 29.39 & -25 34 10.0 & 19.21 & 20 & 0 14 49.86 & -39 20  0.3 & 19.22 \\  
21 & 0 44 47.55 & -25 39 16.4 & 19.23 & 21 & 0 14 49.49 & -39 35  7.6 & 19.22 \\  
22 & 0 46 52.64 & -25 23 35.8 & 19.25 & 22 & 0 12 55.67 & -39 46 29.9 & 19.26 \\  
23 & 0 43 54.16 & -25 30 53.1 & 19.27 & 23 & 0 13  4.33 & -39 43 43.7 & 19.30 \\  
24 & 0 43 40.58 & -25 14 40.3 & 19.33 & 24 & 0 12 42.38 & -39 6 13.5 & 19.32 \\   
25 & 0 43 43.70 & -25 19 50.2 & 19.41 & 25 & 0 14 41.43 & -39 35 14.1 & 19.33 \\  
26 & 0 44  2.69 & -25 11 29.1 & 19.42 & 26 & 0 14 40.96 & -39 23 13.0 & 19.34 \\  
27 & 0 44 24.04 & -25 19 48.0 & 19.43 & 27 & 0 14  7.81 & -39 14 38.9 & 19.37 \\  
28 & 0 45  8.48 & -25 34 55.7 & 19.43 & 28 & 0 13  1.59 & -39 52 22.8 & 19.40 \\  
29 & 0 44 47.84 & -25 29 55.3 & 19.46 & 29 & 0 14  6.12 & -39 28 15.2 & 19.42 \\  
30 & 0 46 18.62 & -25 12 10.8 & 19.54 & 30 & 0 10 35.36 & -39 19 12.2 & 19.44 \\  
31 & 0 46  7.82 & -25 31 25.5 & 19.54 & 31 & 0 11 30.20 & -39 24 56.6 & 19.44 \\  
32 & 0 45 20.82 & -25 35  0.5 & 19.56 & 32 & 0 11 45.84 & -39 27 33.7 & 19.49 \\  
33 & 0 44 36.24 & -25 13  3.8 & 19.58 & 33 & 0 13  1.89 & -39 31 30.6 & 19.51 \\  
34 & 0 46 27.09 & -25 17 27.8 & 19.62 & 34 & 0 13 28.69 & -39 40 53.4 & 19.52 \\  
35 & 0 43 56.76 & -25 26  6.5 & 19.73 & 35 & 0 13 12.97 & -39 48 35.3 & 19.52 \\  
36 & 0 46 20.53 & -25 7 30.0 & 19.74 & 36 & 0 13 37.10 & -39 41  1.9 & 19.53 \\ 
37 & 0 44 26.05 & -25 33 27.7 & 19.78 & 37 & 0 12  1.03 & -39 38 13.4 & 19.56 \\  
38 & 0 44 23.19 & -25 34 27.1 & 19.78 & 38 & 0 13 10.13 & -39 29 22.5 & 19.59 \\  
39 & 0 45  5.08 & -25 41  4.9 & 19.78 & 39 & 0 12 37.46 & -39 40 27.5 & 19.59 \\  
40 & 0 45 26.88 & -25 14  3.7 & 19.80 & 40 & 0 12 56.36 & -39 11  5.5 & 19.61 \\  
41 & 0 46 38.15 & -25 38  5.1 & 19.84 & 41 & 0 12 24.07 & -39 21 38.4 & 19.66 \\  
42 & 0 45 17.95 & -25 7 18.2 & 19.85 & 42 & 0 13 16.92 & -39 30 21.8 & 19.69 \\ 
43 & 0 44 21.09 & -25 9 56.4 & 19.85 & 43 & 0 11 15.90 & -39 41 46.8 & 19.72 \\ 
44 & 0 44 13.97 & -25 16 45.2 & 19.89 & 44 & 0 13 27.56 & -39 31  8.2 & 19.74 \\  
45 & 0 46 11.34 & -25 34 21.7 & 19.89 & 45 & 0 13 48.31 & -39 32 42.0 & 19.77 \\  
46 & 0 43 36.63 & -25 25 59.0 & 19.92 & 46 & 0 11 44.80 & -39 52 46.7 & 19.77 \\  
47 & 0 45 37.20 & -25 30  0.9 & 19.92 & 47 & 0 13 43.24 & -39 16 23.9 & 19.82 \\  
48 & 0 45 31.88 & -25 13 15.9 & 19.93 & 48 & 0 13  7.18 & -39 30 30.2 & 19.82 \\  
49 & 0 46 22.68 & -25 41 22.3 & 19.95 & 49 & 0 12 55.83 & -39 54 42.5 & 19.86 \\  
50 & 0 44 36.63 & -25 35 13.0 & 19.96 & 50 & 0 11 54.33 & -39 6 26.1 & 19.87 \\   
51 & 0 45 36.56 & -25 25 22.5 & 19.97 & 51 & 0 14 43.92 & -39 30 53.4 & 19.90 \\  
52 & 0 43  2.99 & -25 29  3.3 & 19.97 & 52 & 0 12 23.58 & -39 26 38.9 & 19.95 \\  
53 & 0 46 20.52 & -25 18 24.7 & 19.99 & 53 & 0 12  3.73 & -39 42 27.3 & 19.98 \\  
54 & 0 45 32.61 & -25 19 12.7 & 20.00 & 54 & 0 10 41.10 & -39 15 33.7 & 19.99 \\  
55 & 0 45  7.70 & -25 7 49.0 & 20.02 &  55 & 0 10 13.53 & -39 21 36.1 & 19.99 \\ 
56 & 0 46 30.20 & -25 18 57.3 & 20.02 & 56 & 0 12  3.47 & -39 46 56.0 & 20.04 \\  
57 & 0 43 40.49 & -25 18 60.0 & 20.03 & 57 & 0 11  6.42 & -39 36 20.8 & 20.09 \\  
58 & 0 44 12.65 & -25 25  0.7 & 20.03 & 58 & 0 14 21.57 & -39 39  0.1 & 20.14 \\  
59 & 0 45 50.94 & -25 26 24.1 & 20.03 & 59 & 0 14  3.89 & -39 40 26.1 & 20.14 \\  
60 & 0 45 20.69 & -25 24 50.2 & 20.04 & 60 & 0 12 46.07 & -39 46 11.6 & 20.16 \\  
61 & 0 45 32.76 & -25 26 14.9 & 20.04 & 61 & 0 13 28.03 & -39 34 34.8 & 20.17 \\  
62 & 0 45  3.07 & -25 26  6.2 & 20.05 & 62 & 0 11 30.88 & -39 6  5.7 & 20.20 \\   
63 & 0 44 36.71 & -25 28 30.0 & 20.05 & 63 & 0 13 13.77 & -39 54 20.3 & 20.20 \\  

\hline
\end{tabular}
\end{table*}

\begin{table*}
\contcaption{}
\centering
\begin{tabular}{llllllll}

\hline
NGC~253 & & & & NGC~55 & & & \\
ID & $\alpha$(1950) & $\delta$(1950) & $B$ & ID & $\alpha$(1950) & $\delta$(1950) & $B$ \\
\hline

64 & 0 43 14.65 & -25 21 54.9 & 20.09 & 64 & 0 11 36.53 & -39 45 59.4 & 20.21 \\  
65 & 0 45  4.34 & -25 23 34.2 & 20.13 & 65 & 0 11 57.43 & -39 28 54.7 & 20.22 \\  
66 & 0 45 37.23 & -25 23 39.7 & 20.17 & 66 & 0 11 23.35 & -39 55 17.1 & 20.23 \\  
67 & 0 44 40.80 & -25 11  6.0 & 20.19 & 67 & 0 11 16.33 & -39 20 21.0 & 20.25 \\  
68 & 0 45 37.02 & -25 25 54.8 & 20.20 & 68 & 0 12 36.07 & -39 31  8.1 & 20.25 \\  
69 & 0 44 32.62 & -25 36 32.1 & 20.21 & 69 & 0 13 35.26 & -39 16 50.7 & 20.26 \\  
70 & 0 45 34.90 & -25 12 47.7 & 20.25 & 70 & 0 12 58.36 & -39 6 24.0 & 20.27 \\   
71 & 0 44 55.68 & -25 27 17.0 & 20.25 & 71 & 0 12 45.49 & -39 32  5.2 & 20.27 \\  
72 & 0 44  4.62 & -25 33 56.6 & 20.26 & 72 & 0 13 43.07 & -39 0 34.6 & 20.30 \\   
73 & 0 46 32.61 & -25 24 17.2 & 20.27 & 73 & 0 12 44.11 & -39 5 14.9 & 20.35 \\   
74 & 0 44 41.69 & -25 41 10.1 & 20.27 & 74 & 0 13 47.95 & -39 21 47.3 & 20.37 \\  
75 & 0 47  2.19 & -25 16 21.5 & 20.29 & 75 & 0 12  2.75 & -39 29 46.9 & 20.38 \\  
76 & 0 45 24.03 & -25 27 38.9 & 20.30 & 76 & 0 12 47.66 & -39 32  1.9 & 20.40 \\  
77 & 0 45 33.33 & -25 19 35.5 & 20.31 & 77 & 0 13 26.69 & -39 51 25.4 & 20.40 \\  
78 & 0 46 56.06 & -25 15 35.7 & 20.32 & 78 & 0 14 20.98 & -39 39 19.8 & 20.43 \\  
79 & 0 46 19.62 & -25 16 33.5 & 20.34 & 79 & 0 13  3.11 & -39 23 17.3 & 20.45 \\  
80 & 0 46 14.77 & -25 26 23.3 & 20.34 & 80 & 0 14 54.78 & -39 38 33.2 & 20.46 \\  
81 & 0 46 25.46 & -25 29 25.5 & 20.36 & 81 & 0 13 53.54 & -39 14 38.0 & 20.48 \\  
82 & 0 46 20.15 & -25 17 50.8 & 20.38 & 82 & 0 13 39.53 & -39 35 53.9 & 20.48 \\  
83 & 0 44 28.43 & -25 28 57.1 & 20.38 & 83 & 0 13 49.87 & -39 10 44.9 & 20.49 \\  
84 & 0 43 22.61 & -25 21 44.0 & 20.40 & 84 & 0 13 25.27 & -39 52 10.6 & 20.49 \\  
85 & 0 44 36.96 & -25 13  5.7 & 20.41 & & & &  \\
86 & 0 44 31.90 & -25 41 39.1 & 20.41 & & & &  \\
87 & 0 44 25.59 & -25 41 47.8 & 20.47 & & & &  \\
88 & 0 44 50.77 & -25 8 33.0 & 20.49 & & & &  \\
89 & 0 43 29.80 & -25 17  5.3 & 20.49 & & & &  \\
90 & 0 45 45.57 & -25 13  2.6 & 20.50 & & & &  \\
91 & 0 45  1.99 & -25 23 58.9 & 20.50 & & & &  \\

\hline
\end{tabular}
\end{table*}

An idea of the contamination in the new globular cluster samples by 
background galaxies can be gained from a scaling of the relative numbers 
of spectroscopically identified objects within the defined parameter space. 
For the NGC~253 sample, there are 7 globular clusters and 7 background 
galaxies. 
Assuming that there are no foreground stars in the new cluster samples, 
then we expect to obtain $\sim$ 45 clusters in the new NGC~253 sample. 
In the case of NGC~55, there are 6 galaxies and one globular cluster 
within the sample limits, giving an expected $\sim$ 14 globular clusters 
in our sample.

The specific frequencies of the two galaxies may be calculated from an 
estimate of their total cluster populations.
The turn over in the Galactic globular cluster luminosity function (GCLF) 
occurs at $M_{V}$ $=$ -7.6, $\sigma = $ 1.2 \cite{harris99}.
At a distance of 2.5~Mpc for the Sculptor group, this corresponds to 
$B = 19.9~(B-V=0.5)$. Therefore the cluster candidate selection cut 
reaches some $\sim$ 0.6 magnitudes past the turnover. 
This implies a total cluster population of approximately 60 for NGC~253, 
which, with $M_{V}$ $=$ -20.0, yields a specific frequency, 
$S_{\rm N}$ $=$ 0.6.
Applying the same arguments for NGC~55, an SB(s)m galaxy  
with  $M_{V}$ $=$ -19.5 yields an expected total cluster population of
approximately 20 and $S_{\rm N}$ $\sim$ 0.3.

\section{Conclusions}

We have identified 14 globular clusters in the spiral galaxy NGC~253, and 
one possible globular cluster belonging to NGC~55. Using digitized plate 
scans combined with spectroscopically identified stars, galaxies and globular 
clusters has allowed us to create new samples for the two galaxies which 
we expect will contain many new globular clusters.

Automated image searching techniques provide an efficient and, 
more importantly, quantitative way of identifying globular clusters 
from digitized wide--field photographic plates. Nevertheless, 
locating a small number of objects which can exhibit a range properties 
from within an initially large dataset is not straightforward, 
and automated searches \emph{still} need to be supplemented by visual 
examination to minimize contamination from other sources. 
The level of contamination in the Liller \& Alcaino (1983a,b) 
and Blecha (1986) samples indicates the difficulty faced when 
undertaking searches of this type based on geometrical and/or photometric 
properties, especially in relatively poor cluster systems.

\newpage

\end{document}